\documentclass[prl,english,superscriptaddress, 
twocolumn,preprintnumbers,amsmath,amssymb]{revtex4-2}

\usepackage[T1]{fontenc}
\usepackage[latin9]{inputenc}
\usepackage{color}
\usepackage{xcolor}
\usepackage{babel}
\usepackage{amsmath}
\usepackage{amssymb}
\usepackage[pdfusetitle,
 bookmarks=true,bookmarksnumbered=false,bookmarksopen=false,
 breaklinks=false,pdfborder={0 0 0},pdfborderstyle={},backref=false,colorlinks=true]
 {hyperref}
\hypersetup{
 linkcolor=blue, citecolor=cyan, urlcolor=blue}

\makeatletter

\pdfpageheight\paperheight
\pdfpagewidth\paperwidth

\allowdisplaybreaks[2]

\makeatother

\begin{document}
\title{Generalized fluctuation-dissipation theorem and Einstein relation \\ 
in rotating equilibrium}
\author{Shuo Fang}
\email{fangshuo@mail.ustc.edu.cn}

\affiliation{Department of Modern Physics, University of Science and Technology
of China, Anhui 230026, China}
\author{Shi Pu}
\email{shipu@ustc.edu.cn}

\affiliation{Department of Modern Physics, University of Science and Technology
of China, Anhui 230026, China}
\affiliation{Southern Center for Nuclear-Science Theory (SCNT), Institute of Modern
Physics, Chinese Academy of Sciences, Huizhou 516000, Guangdong Province,
China}
\begin{abstract}
We derive a generalized fluctuation--dissipation theorem (FDT) for vector fields in rotating thermal
equilibrium from an exact phase-space Kubo--Martin--Schwinger (KMS) relation.
The resulting FDT contains rotation-induced tensorial contributions beyond a scalar thermal factor.
In the local transport limit, we obtain a model-independent generalized Einstein relation with rotation in which,
through second order in thermal vorticity, symmetric momentum diffusion depends on low-frequency spectral
information beyond that encoded in the rotation-dependent dissipative drag.
Remarkably, we find a new Einstein-type relation linking zero modes in the spectral function to the nonrotating
drag coefficient.
These zero modes underlie a new mechanism for the orbital polarization of heavy quarkonium.
The same FDT constraint yields a modified detailed-balance relation whose first-order vortical correction
is governed by transition polarization, without explicit dependence on the bath model.
Our findings establish microscopic equilibrium constraints for studying rotating quantum matter in various fields.
\end{abstract}
\maketitle

\paragraph{Introduction: }

Thermal equilibrium constrains correlations and responses in many-body systems without requiring a complete solution of their microscopic dynamics. In quantum systems, the Kubo--Martin--Schwinger (KMS) relation connects thermal correlation functions with different operator orderings through imaginary-time evolution \citep{Kubo:1957mj,Martin:1959jp} and, in the algebraic formulation of quantum field theory, characterizes thermal equilibrium at a given inverse temperature \citep{Haag:1996hvx}. An important consequence of this relation is the fluctuation-dissipation theorem (FDT), which connects spontaneous equilibrium fluctuations to the dissipative response induced by weak external perturbations \citep{Callen:1951vq,Kubo:1957mj}. In the transport limit, this connection yields the Einstein relation, which fixes the relative strengths of momentum diffusion and drag for nonrelativistic Brownian motion in an isotropic, nonrotating medium \citep{Einstein:1905rjn,Svetitsky:1987gq}. The same equilibrium constraints on thermal correlation functions also underlie detailed balance between forward and reverse quantum transitions \cite{Maxwell:1867,Einstein:1917zz}. We refer the reader to the modern textbooks~\citep{Altland:2010lla,Calzetta:2008iqa} for detailed discussions.

These equilibrium relations raise a further question for thermal states constrained by the conservation of total angular momentum in addition to energy and particle number.
Rotation plays an important role in a wide range of physical systems.
Examples include Barnett magnetization \citep{RevModPhys.7.129}, spin Hall currents generated by fluid vorticity \citep{Takahashi2016}, and quantized vortices
in rotating Bose--Einstein condensates \citep{RevModPhys.81.647}.
Rotation also affects massive-star structure and evolution \citep{RevModPhys.84.25} and can induce phase
transitions in models of strongly interacting matter \citep{Chen:2015hfc,Jiang:2016wvv}.
Spin, the intrinsic contribution to total angular momentum, is central in areas from  quantum materials
\citep{Zutic:2004zz}  to nucleon structure \citep{Aidala:2012mv}.
Meanwhile, recent observations of spin polarization induced by initial orbital angular momentum~\cite{Liang:2004xn,Liang:2004ph} in relativistic heavy-ion collisions provide a new
opportunity to study the interplay between orbital and spin angular momentum and indicate that the
quark--gluon plasma produced in these collisions is the
``\textit{most vortical fluid}''~\cite{STAR:2017ckg,Becattini:2016gvu}.
See also the recent reviews~\cite{Hidaka:2022dmn,Gao:2020vbh,Becattini:2024uha} and references therein.
These observations motivate studying the KMS relation and FDT in rotating  systems.

Although the KMS relation has been studied separately in rotating \citep{Alicki:2017bgh,Salvio:2026ewl} and accelerating systems \citep{Crispino:2007eb,Ambrus:2023smm}, a unified KMS formulation incorporating both effects remains to be established. 
For spin transport, it is also important to formulate the rotating KMS relation in phase space for spinful particles with explicit spin dependence.
Recently, the FDT in rotating equilibrium has been studied in Ref.~\citep{Kush:2026zjq}, but the corresponding Einstein relation was imposed as an additional input rather than derived from the rotating FDT. Earlier studies of Brownian transport have related  concentration diffusion to rotation-dependent mobility through a presumed Einstein relation \citep{PhysRevLett.61.1442} and related fluctuating-force correlations to friction \citep{gotoh1990brownian}.
Against these background, a model-independent formulation of these equilibrium constraints 
without a weak-coupling expansion is still needed.
Such a formulation would constrain spin transport in media such as the quark--gluon plasma without
explicitly solving the interacting field equations.
In particular, we ask whether the angular-momentum conservation constraint
\textit{changes the Einstein relation itself by making momentum diffusion depend on low-frequency response
information beyond dissipative drag}.


To address these questions, we derive an exact phase-space KMS relation for vector fields incorporating rotation
and acceleration and obtain a generalized FDT with rotation-induced tensorial contributions
beyond a scalar thermal factor. Applying this FDT in the local transport limit, we obtain a model-independent generalized Einstein relation
with rotation in which, through second order in thermal vorticity, symmetric momentum diffusion depends
on low-frequency spectral information beyond that encoded in the rotation-dependent dissipative drag. 
We also identify rotation-induced zero modes in the spectral function and derive a new Einstein-type relation
linking these zero modes to the nonrotating drag coefficient. 
Such zero modes can further lead to orbital polarization of heavy quarkonium.  The same FDT constraint yields a modified detailed-balance relation coupling vorticity to transition polarization. To the best of our knowledge, neither the generalized Einstein relation nor the new Einstein-type relation for spectral zero modes has been reported previously. These results provide microscopic equilibrium constraints for a broad range of rotating quantum systems.
Throughout this work, we use the mostly-minus
Minkowski metric and the (anti-)symmetrization conventions
$a_{[\mu \nu]}=a_{\mu \nu}-a_{\nu \mu}$ and $a_{(\mu \nu)}=a_{\mu \nu}+a_{\nu \mu}$. 


\vspace{1em}
\paragraph*{Conventional KMS relation and FDT:}
We consider a many-body system in grand-canonical equilibrium at temperature $T$ and chemical potential $\mu$. We use bosonic vector fields unless stated otherwise.
For two Heisenberg-picture operators $A(t)$ and $B(t^{\prime})$, cyclicity of the trace and Heisenberg evolution give $\left\langle A(t)B(t^{\prime})\right\rangle \equiv\textrm{Tr}[\varrho A(t)B(t^{\prime})] = e^{-\mu/T}
\left\langle B(t^{\prime})A(t+iT^{-1})\right\rangle $,
where $\varrho=Z^{-1}\exp[-(H-\mu N)/T]$ is the density operator, with $Z$ the partition function, $H$ the Hamiltonian, and $N$ the particle-number operator. 
This is the \textit{KMS relation} \citep{Haag:1996hvx}, which connects the two operator orderings through the imaginary-time translation $t\rightarrow t+iT^{-1}$ together with the corresponding chemical-potential factor.

For a complex vector field $V_{\rho}(x)$ carrying one unit of this charge, the KMS condition relates the two Wightman functions \citep{Kubo:1957mj,Martin:1959jp}
$G_{\rho\sigma}^{<}(x;x^{\prime})=\langle V_{\sigma}^{\dagger}(x^{\prime})V_{\rho}(x)\rangle$
and
$G_{\rho\sigma}^{>}(x;x^{\prime})=\langle V_{\rho}(x)V_{\sigma}^{\dagger}(x^{\prime})\rangle$
according to
\begin{eqnarray}
G_{\rho\sigma}^{>}(x;x^{\prime})
&=&
e^{-\mu/T}G_{\rho\sigma}^{<}(t+iT^{-1},\boldsymbol{x};x^{\prime}).
\label{eq:G>andG<}
\end{eqnarray}
Here, $\rho$ and $\sigma$ are Lorentz indices, and additional internal indices are suppressed. 
Eq.~(\ref{eq:G>andG<}) reduces the number of independent thermal two-point functions and leads directly to the FDT. Suppressing the Lorentz indices, we define the bosonic spectral function
$\rho(x;x^{\prime})=G^{>}(x;x^{\prime})-G^{<}(x;x^{\prime})$
and the statistical function
$G^{F}(x;x^{\prime})=\frac{1}{2}\left[G^{>}(x;x^{\prime})+G^{<}(x;x^{\prime})\right]$.
In a translationally invariant equilibrium state, both depend only on $x-x^{\prime}$.
By using Eq.~(\ref{eq:G>andG<}), we obtain after Fourier transformation
\begin{eqnarray}
G^{F}(p)
&=&
\frac{1}{2}
\coth\left[\frac{p_{0}-\mu}{2T}\right]
\rho(p).
\label{eq:FDT-bosons}
\end{eqnarray}
In the Schwinger-Keldysh formalism \citep{Altland:2010lla,Calzetta:2008iqa}, $G^{F}$ encodes statistical fluctuations, whereas $\rho$ characterizes the spectrum and the absorptive part of the linear response.
Therefore, Eq.~(\ref{eq:FDT-bosons}) relates equilibrium fluctuations to dissipative responses through a scalar Bose thermal factor
and is known as the \textit{fluctuation-dissipation theorem} for bosons \citep{Callen:1951vq,Kubo:1957mj}.

\vspace{1em}
\paragraph*{Generalized KMS relation in rotating equilibrium:}
We next consider global equilibrium in a relativistic system with conserved total angular momentum. Maximizing the von Neumann entropy subject to energy-momentum, particle-number, and total-angular-momentum conservation gives the density operator $\varrho_J  =  Z_J^{-1}\exp\left[-b_{\mu}P^{\mu}+\alpha N+\varpi_{\mu\nu}J^{\mu\nu}\right]$ \cite{zubarev1974nonequilibrium},
with constant Lagrange multipliers $\{b_{\mu},\alpha,\varpi_{\mu\nu}\}$.
Here $P^{\mu}$ and $J^{\mu\nu}$ are the generators of translations and Lorentz transformations satisfying the Poincar\'e algebra, while $N$ is the particle-number generator commuting with them. 
Later we will show that, at global equilibrium, $b^\mu$ is the constant part of the inverse-temperature four-vector,
$\alpha=\mu/T$, and the anti-symmetric tensor $\varpi_{\mu\nu}$ is the intensive parameter
conjugate to total angular momentum.
As in the conventional case, cyclicity of the trace gives, for two operators $A(x)$ and $B(x^\prime)$, 
\begingroup
\setlength{\abovedisplayskip}{5pt}
\setlength{\abovedisplayshortskip}{2pt}
\setlength{\belowdisplayskip}{4pt}
\setlength{\belowdisplayshortskip}{4pt}
\begin{eqnarray}
\left\langle \frac{}{} A(x)B(x^{\prime})\right\rangle_J =
e^{-\mu/T} \left\langle B(x^{\prime})\tilde{A}(x;b,\varpi)\right\rangle_J . 
\label{eq:G_KMS} 
\end{eqnarray} 
\endgroup
Here $\langle\cdots\rangle_J$ denotes averaging with $\varrho_J$,
and $\tilde{A}(x;b,\varpi)\equiv e^{-C(P,J)}A(x)e^{+C(P,J)}$,
with $C(P,J)=b \cdot P-\varpi_{\mu\nu}J^{\mu\nu}$.
Eq.~(\ref{eq:G_KMS}) is the \textit{generalized KMS relation} in rotating equilibrium. Compared with the conventional relation, 
$\tilde{A}$ is no longer obtained by a simple imaginary-time translation. Instead, it incorporates both spacetime translations and Lorentz transformations, including boosts and rotations.
For the bosonic Wightman functions $G$, Eq.~(\ref{eq:G_KMS}), together with the Poincar\'e algebra
and the transformation law of vector fields, gives
\begin{eqnarray}
G^{>}(x;x^{\prime}) & = & e^{-\mu/T}e^{-R}e^{i\beta(x)\cdot \partial_{x}}G^{<}(x;x^{\prime}).
\label{eq:G_S>}
\end{eqnarray}
Here $R_{\alpha\beta}=\frac{1}{2}\varpi_{\mu\nu}\Sigma^{\mu\nu}_{\alpha\beta}$,
where $\Sigma^{\mu\nu}_{\alpha\beta}=i\eta^{[\mu}_\alpha \eta^{\nu]}_\beta$
are the Lorentz generators of the vector field.  The vector $\beta_\mu(x)=b_{\mu}+2\varpi_{\mu\nu}x^{\nu}$ will later be identified with the inverse-temperature four-vector at global equilibrium. 
The factors $e^{-R}$ and $e^{i\beta\cdot\partial_x}$ encode the spin and spacetime parts
of the generalized transformation, respectively.
This structure makes the spin dependence of equilibrium correlations explicit and provides the starting point for the generalized FDT derived below.

\vspace{1em}
\paragraph*{Phase-space KMS relation:}
We next need to recast Eq.~(\ref{eq:G_S>}) in phase space. 
In rotating global equilibrium, the angular-momentum  conservation constraint selects a preferred axis.
Such a state is generally not invariant under arbitrary spatial  rotations and translations. Its two-point functions therefore depend separately on the center coordinate $X\equiv(x+x^\prime)/2$ and the relative coordinate $x-x^\prime$. A momentum-space description that discards the $X$ dependence is insufficient.

We accordingly define the Wigner transform of a Wightman function $G$ by Fourier transforming the relative coordinate 
\footnote{Rigid rotation in a relativistic system is restricted to a finite transverse region by causality
\cite{Vilenkin:1979ui}.
For slow rotation, we focus on the bulk near the rotation axis, far from this boundary,
and use the continuum Wigner representation, neglecting boundary corrections.},
$G(p,X)=\int \mathrm{d}^4y\,e^{ip\cdot y}G(X+y/2;X-y/2)$.
Here $p$ is the four-momentum conjugate to $y$ and can be interpreted as the kinetic momentum in the quasiparticle limit.
Using the Poincar\'e algebra and the Wigner transformation, we obtain the exact phase-space KMS relation,
\begingroup
\setlength{\abovedisplayskip}{5pt}
\setlength{\abovedisplayshortskip}{2pt}
\setlength{\belowdisplayskip}{4pt}
\setlength{\belowdisplayshortskip}{4pt}
\begin{align}
e^{\mu/T-p\cdot\tilde{b}}\,
G^>\!\left(p,X-\frac{i\tilde{b}}{2}\right)
={}&e^{-R}e^{-\varpi_{\rho\sigma}\mathcal{L}^{\rho\sigma}}
\nonumber\\
&\times G^<\!\left(e^{i\varpi}p,e^{i\varpi}X\right), \label{eq:New_G}
\end{align}
\endgroup
where $\mathcal{L}^{\rho\sigma}=X^{[\rho} p^{\sigma]}+\partial_p^{[\rho}\partial_X^{\sigma]}/4$ 
represents the orbital Lorentz transformation in phase space,
combining the classical orbital term with a mixed derivative term generated by the Wigner transform.
Thus, $R$ encodes the intrinsic-spin transformation, while the orbital effect of rotation enters in two ways, through $\mathcal{L}$ and through the rotation of the phase-space coordinates $p$ and $X$.
Here $\tilde{b}=(1-e^{-2i\varpi})b/(2i\varpi)$ is the effective translation vector obtained from the Poincar\'e factorization \citep{Becattini:2020qol}, where the matrix function is defined by its power-series expansion.

\vspace{1em}
\paragraph*{Generalized fluctuation-dissipation relation:}
We derive the generalized FDT by introducing the spectral function and the statistical function in phase space. 
Although Eq.~(\ref{eq:New_G}) is exact, its full rotational dependence obscures the underlying physics.
We first resum the rotational corrections to the thermal occupation factor into a generalized distribution
$n(p,X)=[e^{\beta(X)\cdot p-\mu/T}-1]^{-1}$.
Here $\beta_\mu(X)=b_\mu+2\varpi_{\mu\nu}X^\nu$ follows from the generalized KMS relation.
It satisfies the Killing condition $\partial_{(\mu}\beta_{\nu)}=0$ at global equilibrium \citep{Becattini:2014yxa}.
The local temperature $T(X)=1/\sqrt{\beta^\mu\beta_{\mu}}$ and $\mu$ are now position dependent,
while their ratio $\mu/T$ remains constant \citep{Landau:1980mil}.
These conditions agree with those obtained from the Boltzmann equation at global equilibrium
\cite{groot1980relativistic}.
These relations identify $\beta_\mu$ and $\mu$ as the inverse-temperature four-vector and chemical potential,
respectively.
The explicit form of $\beta_\mu(X)$ further identifies $2\varpi_{\mu\nu}$ as the thermal vorticity tensor
$-\frac{1}{2}\partial_{[\mu}\beta_{\nu]}$ \citep{Becattini:2014yxa}.

We then expand the remaining contributions in $\varpi_{\mu\nu}$.
To linear order in $\varpi_{\mu\nu}$, the generalized fluctuation-dissipation relation is 
\footnote{We also note that related FDTs have recently been derived for the rotating stochastic Landau model using the Martin--Siggia--Rose path-integral formalism~\cite{Kush:2026zjq}.
However, the Einstein relation there is imposed as an additional input to match these relations to the high-temperature FDT. Besides, the KMS relation used there neglects the intrinsic spin degrees of freedom, which generate
the second term on the right-hand side of Eq.~(\ref{eq:FDT_NEW}).},\vspace{-0.4em}
\begin{align}
G_{\mu\nu}^{F}(p,X)
=&
\frac{1}{2}\coth\!\left[\frac{\beta(X)\cdot p-\mu/T}{2}\right]\rho_{\mu\nu}
\nonumber\\
&+\frac{i}{4}\,\operatorname{csch}^{2}\left[\frac{\beta(X)\cdot p-\mu/T}{2}\right]
\{\varpi,\rho\}_{\mu\nu} . \label{eq:FDT_NEW}
\end{align}
Here $\{\varpi,\rho\}_{\mu\nu}\equiv\varpi_{\mu\alpha}\rho^{\alpha}_{\;\;\nu}+\rho_{\mu}^{\;\alpha}\varpi_{\alpha\nu}$ denotes the symmetrized product of $\varpi$ and $\rho$ in Lorentz-index space. 
Eq.~(\ref{eq:FDT_NEW}) relates equilibrium fluctuations to the spectral function through a scalar thermal factor
and an additional rotation-induced tensorial contribution involving $\varpi_{\mu\nu}$ and $\rho$.
The relation requires neither a specific interaction model nor a weak-coupling expansion
and is therefore nonperturbative in the interaction strength.

\vspace{1em}
\paragraph*{Einstein relation in rotating equilibrium:}
We now turn to the transport implications of the generalized FDT.
For a nonrelativistic particle of mass $M$ described by Langevin dynamics in an isotropic, nonrotating medium, the conventional FDT yields the Einstein relation $\kappa=2MT\eta_{{\rm D}}$ between the momentum-diffusion coefficient $\kappa$
and the drag coefficient $\eta_{{\rm D}}$ \citep{Einstein:1905rjn,Svetitsky:1987gq}.
We now use the generalized FDT to determine how rotation modifies this relation.

We start with a heavy nonrelativistic particle obeying the Langevin equation
$d p^{i}/dt= F^i-\eta_{{\rm D}}^{ij}p^{j}+\xi^{i}(t)$,
where $F^i$ is the conservative force and $\xi^{i}(t)$ is the stochastic force exerted by the medium \citep{gardiner2000quantum}. The Coriolis force and possible conservative forces $\boldsymbol{F}$ can induce a position-dependent momentum shift in the diffusion kernel, but do not directly contribute to $\eta_{\rm D}^{ij}$.
For simplicity, we consider the system near the rotation axis and neglect explicit position dependence in the dissipative dynamics.
Within this local approximation, rotational effects on dissipative Langevin dynamics enter only through
the medium correlation functions. We further specialize to pure rotation in the local rest frame, retaining only $\varpi_{ij}$.
The correlators then depend only on the relative time, with $E$ denoting the conjugate energy.
The transport coefficients follow from the Wightman correlators $G^{>}$ and $G^{<}$,
now defined for the microscopic force operator.
Linear response gives the drag coefficient through
$2M\eta_{{\rm D}}^{ij}=\lim_{E\to0}\rho^{ij}(E)/E$ \citep{Kapusta:2006pm}.
The transport limit $E\to0$ probes  the macroscopic dynamics with time scales much longer than microscopic relaxation times.

\vspace{0.3em}
\noindent
\textbf{Spectral and statistical functions with rotation}
---
To identify how rotation enters, we focus on the low-frequency structure of the spectral function $\rho(E)$. Since the noise force $\xi(t)$ is Hermitian, the Wigner transform of the spectral function $\rho_{ij}(t)=\langle [\xi_i(t),\xi_j(0)] \rangle$ satisfies $\rho_{ij}^\dagger(E) = \rho_{ji}(E)$. 
Thus, at small $E$, we parameterize the spectral function as
\footnote{The spectral function may contain nonanalytic contributions,
such as terms proportional to fractional powers of $E$ \citep{Burnier:2010rp}.
However, these terms do not contribute to the final Einstein relation
once the finiteness constraint on $\kappa_{ij}$ is imposed.}
\begin{eqnarray}
\rho^{ij}(E) & = & i \gamma_0^{ij}+2E\gamma^{ij}_{1}+iE^{2}\gamma_{2}^{ij}+E^3 \gamma^{ij}_3+\mathcal{O}(E^4) . \label{eq:rho_E}
\end{eqnarray}
Here, the symmetric tensors $\gamma^{ij}_{1,3}$ and the antisymmetric tensors $\gamma_{0,2}^{ij}$ are all real. The zero mode $i\gamma_0^{ij}$ is usually absent without rotation
but is allowed by symmetry constraints. 
For the purely spatial thermal vorticity $2\varpi_{ij}$ considered here, microscopic time-reversal symmetry,
under which $\varpi_{ij}\rightarrow -\varpi_{ij}$, implies the Onsager reciprocity relation
$\rho_{ij}^*(E,\varpi_{ij})=\rho_{ij}(E,-\varpi_{ij})$ \cite{Onsager:1931kxm}.
Consequently, $\gamma^{ij}_{1,3}$ are even in $\varpi_{ij}$, whereas $\gamma_{0,2}^{ij}$ are odd. For small $\varpi_{ij}$, their leading behaviors are $\gamma^{ij}_{1,3}\sim\mathcal{O}(1)$, and $\gamma^{ij}_{0,2}\sim\mathcal{O}(\varpi)$. 
We next consider the fluctuation sector, characterized by the particle momentum-diffusion tensor $\kappa^{ij}$.
In the corresponding Fokker--Planck description,  $\kappa^{ij}$ is given by the noise correlator $\langle\xi^{i}(t)\xi^{j}(t^{\prime})\rangle=\kappa^{ij}\delta(t-t^{\prime})$ and related to the statistical function via $\kappa_{ij}=\lim_{E\to0}G^{F}_{(ij)}(E)/2$ \citep{Moore:2004tg,Casalderrey-Solana:2006fio}. 
For the white noise considered here, the momentum-diffusion tensor is symmetric, $\kappa^{ij}=\kappa^{ji}$. The same argument as for $\rho_{ij}$ gives $G^F_{ij}(E)=G^F_{ji}(-E)$. 

Extending the generalized FDT in Eq.~(\ref{eq:FDT_NEW}) to $\mathcal{O}(\varpi^2)$
and using the transport relations above, we obtain in the transport limit
\vspace{-0.4em}
\begin{eqnarray}
\kappa^{ij} & = & MT\eta_{{\rm D}}^{(ij)}+T^{2}\gamma_{2}^{k(i}\varpi^{j)k}-2T^{3}\gamma_{3}^{l(i}\varpi^{j)k}\varpi^{kl}. \label{eq:Einstein_New}
\end{eqnarray}
In the $\varpi^{ij}\to0$ limit, Eq.~(\ref{eq:Einstein_New}) reduces to the conventional Einstein relation \cite{Einstein:1905rjn}. The second and third terms on the right-hand side of Eq.~(\ref{eq:Einstein_New}) are the leading
rotation-induced corrections and couple the higher-order coefficients $\gamma_{2,3}^{ij}$
in the low-frequency spectral expansion to symmetric momentum fluctuations.
Although $\gamma_{2,3}^{ij}$ depend on the microscopic dynamics,  their tensor structures in Eq.~(\ref{eq:rho_E}) are fixed by equilibrium symmetry constraints.
Eq.~(\ref{eq:Einstein_New}) therefore gives the model-independent form of the leading rotational correction to the Einstein relation. 
By comparison, vorticity corrections in Ref.~\cite{PhysRevLett.61.1442} entered through the spectral function
and affected only the antisymmetric part of $\kappa^{ij}$, and the conventional Einstein relation was used
there in matrix form without an explicit derivation.

\vspace{0.3em}
\noindent
\textbf{Zero modes in the spectral function}
---
The spectral zero mode $i\gamma_0^{ij}$ in Eq.~(\ref{eq:rho_E})
could also contribute to the drag coefficient $\eta_{\text{D}}^{ij}$.
However, requiring the full $\kappa_{ij}$ to remain finite as $E\rightarrow0$
yields a new Einstein-type relation,
\begingroup
\setlength{\abovedisplayskip}{5pt}
\setlength{\abovedisplayshortskip}{2pt}
\setlength{\belowdisplayskip}{4pt}
\setlength{\belowdisplayshortskip}{4pt}
\begin{equation}
  \gamma^{ij}_0=4MT\eta_{\text{D}}\varpi^{ij}, \label{eq:zero_modes}
\end{equation}
\endgroup
where $\eta_{\text{D}}$ is the conventional drag coefficient, i.e., $\eta_{\text{D}}^{ij}\rightarrow\eta_{\text{D}}\delta^{ij}$ in the absence of rotation.  
Under this constraint, the zero-mode correction cancels against other terms
and therefore does not contribute to the drag coefficient up to $\mathcal{O}(\varpi^3)$.
We also emphasize that the relation in Eq.~(\ref{eq:zero_modes}) is nonperturbative in the medium interactions. 
As detailed in the Supplemental Material~\citep{SupplementalMaterial}, we take a hard-thermal-loop calculation in a quantum electrodynamic
plasma as an example and obtain
$\gamma_{0}^{ij}=\varpi^{ij}T^{3}(e^{4}\log e^{-1})/(9\pi)$ together with the corresponding expression for $\eta_{\text{D}}$.

\vspace{1em}
\paragraph{Orbital polarization of heavy quarkonium:}
Heavy-quark dynamics in a medium is an important application of the Langevin equation. We now apply the zero-mode constraint to heavy quarkonium in the highly vortical medium produced in relativistic heavy-ion collisions \cite{STAR:2017ckg} and discuss two physical consequences.
To isolate rotation-induced polarization  effects from spin polarization, we first consider a spin-singlet state. Within the open-quantum-system framework \cite{Akamatsu:2020ypb,Yao:2021lus,Brambilla:2017zei},
the medium's spectral zero modes contribute to the evolution of the quarkonium orbital angular momentum
$\langle L^k \rangle_{J}$ through 
\begin{eqnarray}
\frac{{\rm d}}{{\rm d}t}\langle L^{k}(t)\rangle_{J} & = & \gamma_{0}^{ij}\left[C^{ijk}(t)-\epsilon^{ijk}\frac{\langle\boldsymbol{r}^{2}(t)\rangle}{2(N_{c}^{2}-1)}\right].\label{eq:orbital_polarization}
\end{eqnarray}
where  $C_{ijk}$ is the collision kernel,  $\langle\boldsymbol{r}^{2}(t)\rangle$ is evaluated
with the nonrotating color-octet density operator, and $N_c$ is the number of colors.
 The transport coefficient $\gamma_0^{ij}$ for orbital angular momentum evolution  follows from the zero-mode constraint in Eq.~(\ref{eq:zero_modes}). Eq.~(\ref{eq:orbital_polarization}) thus relates rotation-induced orbital polarization to the drag coefficient governing momentum relaxation. It involves an interplay between strongly coupled QCD interactions
and collective spin-hydrodynamic modes.

A nonzero $L^z$ generated in a rotating medium indicates a population imbalance between states
with opposite orbital angular momentum projections along the $z$ axis.
This orbital polarization leads to a helicity-dependent difference in the photon angular distribution in radiative decays such as $h_c\rightarrow \eta_c+\gamma$.  In the photon helicity frame,  we obtain
$\text{d}(\Delta \Gamma)/(\Gamma \text{d}\cos\theta)=\frac{3}{4} \langle L^z \rangle_J \cos \theta$,
where $\Gamma=\Gamma_+ + \Gamma_-$ and $\Delta \Gamma =\Gamma_+ - \Gamma_-$  with $\Gamma_\pm$ the decay rates into photons of helicity $\pm1$ and $\theta$ the angle between the photon momentum and the $z$ axis.
For spin-triplet heavy quarkonia, we consider hadronic decays such as $\chi_{c1}\to\Lambda\bar{\Lambda}$ \citep{BESIII:2025gof}.
Conservation of total angular momentum, parity, and charge conjugation,
together with the Wigner--Eckart theorem \cite{Sakurai:2011zz},
gives the daughter polarization $P_{\Lambda}^{z} = \langle L^{z}_{\chi_{c1}}\rangle_J$.
The spectral zero modes can therefore contribute to $P_\Lambda$ through Eq.~(\ref{eq:orbital_polarization}).
Here $\Lambda$ polarization is inherited from the orbital polarization of the parent $\chi_{c1}$,
rather than from polarized $s$ quarks through coalescence \citep{Liang:2004ph}.
Further discussion of these two examples is provided in Supplemental Material ~\citep{SupplementalMaterial}.

\vspace{1em}
\paragraph{Detailed balance with rotation: }
The FDT is also closely related to detailed balance \cite{Maxwell:1867,Einstein:1917zz}, which constrains forward and reverse transition rates at equilibrium. 
In a nonrotating system, detailed balance for transitions $|i \rangle \rightleftharpoons |f\rangle$ requires $P_i \Gamma_{i\rightarrow f} = P_f \Gamma_{f\rightarrow i}$, where $P_i$ is the equilibrium occupation probability of $|i\rangle$ and $\Gamma$ denotes the transition rate.
Boltzmann probabilities then give
$\Gamma_{i\rightarrow f}/\Gamma_{f\rightarrow i}=\exp[-(E_f - E_i)/T]$,
with $E_i$ the energy of state $|i\rangle$.
Transition rates are determined by the environment's Wightman functions, which can be expressed in terms of $\rho$   \cite{breuer2002theory}.
The generalized FDT therefore constrains rotational corrections to these rates.

We consider the coupling $\hat{H}_{\text{int}}=\hat{\boldsymbol{J}}\cdot\hat{\boldsymbol{F}}$
between the system current $\hat{J}^{i}$ and the environment's vector force $\hat{F}^{i}$. We describe the reduced dynamics by a Redfield equation in the Born--Markov approximation \citep{breuer2002theory}. The labels $i$ and $f$ cover both discrete atomic levels and continuum states in quantum many-body systems. 
Using Eq.~(\ref{eq:New_G}) to $\mathcal{O}(\varpi)$, we obtain
\begin{eqnarray}
\frac{\Gamma_{i\to f}}{\Gamma_{f\to i}} & = & e^{-(E_f-E_i)/T}\left(1+\frac{2 \boldsymbol{\omega}\cdot \boldsymbol{c}_{if}}{T}\right),
\label{eq:Modified-detailed-balance}
\end{eqnarray}
where $\omega_i\equiv 2 \epsilon_{ijk}T \varpi_{jk}$ is the three-dimensional kinetic vorticity vector, 
$\boldsymbol{c}_{if}\equiv -i \boldsymbol{M}^* \times \boldsymbol{M}/|\boldsymbol{M}|^2$
and $M^{i}=\langle i|\hat{J}^{i}|f\rangle$ is the transition matrix element. Eq.~(\ref{eq:Modified-detailed-balance}) gives the detailed-balance relation in a spinful fluid to first order in vorticity, without explicit dependence on the bath model.

We note that for a quantum many-body bath isotropic at zero rotation, Eq.~(\ref{eq:Modified-detailed-balance}) gives a channel-summed detailed-balance relation that accommodates continuous system energy spectra, with $\boldsymbol{c}_{if}$ encoding the angular-momentum polarization carried by the vector transition matrix element.
Its derivation requires the spin-dependent tensor relation between the vector-force Wightman and spectral
functions, which is not specified by the channel-resolved rate ratio in Ref.~\cite{Alicki:2017bgh} alone.
In Supplemental Material~\citep{SupplementalMaterial}, we illustrate this relation with atomic transitions in a QED cavity
\citep{PhysRevLett.125.241301}, where transition rates can be strongly enhanced.

\vspace{1em}
\paragraph{Summary and outlook: }
In summary, we have derived an exact phase-space KMS relation, Eq.~(\ref{eq:New_G}),
for vector-field Green functions in rotating equilibrium by incorporating total-angular-momentum conservation.
The resulting generalized FDT in Eq.~(\ref{eq:FDT_NEW}) contains rotation-induced tensorial contributions beyond a scalar thermal factor. In the local transport limit, Eq.~(\ref{eq:Einstein_New}) generalizes the Einstein relation
by relating momentum diffusion to low-frequency response beyond the drag tensor.
Through second order in thermal vorticity, the additional terms involve the antisymmetric coefficient
$\gamma_2^{ij}$ and the symmetric coefficient $\gamma_3^{ij}$. Remarkably, we also derive a new Einstein-type relation, nonperturbative in the medium interactions, linking
rotation-induced zero modes in the spectral function to the nonrotating drag coefficient.
These zero modes can induce heavy-quarkonium orbital polarization through a mechanism linked to momentum relaxation. 
The same KMS constraint yields a first-order rotational correction to detailed balance governed
by the transition's angular-momentum polarization, without explicit dependence on the microscopic bath model. 
These results identify how the angular-momentum  conservation constraint shapes equilibrium fluctuations, with consequences for both stochastic transport and microscopic transitions.

Besides, in the quasiparticle limit, our generalized FDT also underlies the equilibrium contributions to the  near-equilibrium Wigner-function polarization.
It can therefore modify spin-dependent bulk quantities, such as the Zilch current~\citep{Chernodub:2018era}, and yield additional nonperturbative corrections to the spin alignment of vector mesons~\citep{Liang:2004xn,STAR:2022fan,Sheng:2022wsy}.
A similar method can also be used to derive the KMS relation and FDT for fermions and study their polarization effects. Such resulting relations provide equilibrium constraints on spin distribution functions in many-body systems. These novel results will be presented in future work.

\vspace{0.5em}
\indent
{\bf Acknowledgments}
We are thankful to Jean-Paul Blaizot, Kenji Fukushima, Di-Lun Yang for helpful discussions. This work is supported in part by the National Key Research and Development Program of China under Contract No. 2022YFA1605500, by the Chinese Academy of Sciences (CAS) under Grants No. YSBR-088 and by National Natural Science Foundation of China (NSFC) under Grants No. 12135011.

\makeatletter
\def\bibsection{%
\par
\onecolumngrid@push
\begingroup
\baselineskip26\p@
\bib@device{\textwidth}{245.5\p@}%
\endgroup
\nobreak
\csname @nobreaktrue\endcsname
\addvspace{19\p@}%
\par
\onecolumngrid@pop
}
\makeatother
\bibliographystyle{apsrev4-1}
\bibliography{qkt-ref}
\onecolumngrid

\clearpage

\onecolumngrid
\setcounter{page}{1}

\begin{center}
{\large\bfseries Supplemental Material for }
{\large\bfseries ``Generalized fluctuation-dissipation theorem and Einstein relation 
in rotating equilibrium''}

\vspace{0.8em}

Shuo Fang$^{1}$ and Shi Pu$^{1,2}$
\vspace{0.4em}

{\itshape
$^{1}$Department of Modern Physics,
University of Science and Technology of China,
Hefei 230026, China \\
\itshape
$^{2}$ Southern Center for Nuclear-Science Theory (SCNT), Institute of Modern
Physics, \\ Chinese Academy of Sciences, Huizhou 516000, Guangdong Province,
China
}

\end{center}

\setcounter{section}{0}
\setcounter{equation}{0}

\renewcommand{\thesection}{S\arabic{section}}
\numberwithin{equation}{section}
\renewcommand{\theequation}{\thesection.\arabic{equation}}

\vspace{0.1cm}

This Supplemental Material expands on the main-text discussion of the generalized KMS, FDT and Einstein relation with rotation. 
In this Supplemental Material, we omit the subscript $J$ of the expectation value $\langle O\rangle_{J}$
unless otherwise stated.

\section{Symmetry and reciprocal properties of vector-field Green functions}\label{sec:symmetry-properties}
\label{sec:Symmetry}
The hermiticity of the field operators and microscopic time-reversal symmetry
impose constraints on the symmetry properties of two-point functions.
We consider the real vector field $\xi^{\mu}(x)$ for illustration. We define its spectral function 
\begin{eqnarray}
\rho^{\mu\nu}(x,x^{\prime}) & = & \langle[\xi^{\mu}(x),\xi^{\nu}(x^{\prime})]\rangle.
\end{eqnarray}
Hermiticity and the antisymmetry of the commutator give $\rho^{\mu\nu}(x,x^{\prime})  =  -\left[\rho^{\mu\nu}(x,x^{\prime})\right]^{\dagger}=-\rho^{\nu\mu}(x^{\prime},x).$
In the Wigner-transformed phase space, these properties give us the following constraints on $\rho^{\mu\nu}(p,X)$,
\begin{eqnarray}
\left[\rho^{\mu\nu}(p,X)\right]^{\dagger} =-\rho^{\mu\nu}(-p,X)=\rho^{\nu\mu}(p,X) ,
\end{eqnarray}
and in terms of its real and imaginary parts, 
\begin{eqnarray}
{\rm Re}\rho^{\mu\nu}(p,X) & = & -{\rm Re}\rho^{\mu\nu}(-p,X)={\rm Re}\rho^{\nu\mu}(p,X),\quad\quad
{\rm Im}\rho^{\mu\nu}(p,X)  =  {\rm Im}\rho^{\mu\nu}(-p,X)=-{\rm Im}\rho^{\nu\mu}(p,X).
\end{eqnarray}
In other words, the real part of the spectral function is \textit{symmetric} in $\mu,\nu$
and odd under $p\to-p$, whereas its imaginary part is \textit{antisymmetric} in $\mu,\nu$
and even under $p\to-p$.

Another important property of the spectral function is the constraint
from time-reversal symmetry. For this analysis, we restrict to pure rotation in the rest frame, retaining only $\varpi_{ij}$. The antiunitary time-reversal operator $\Theta$ relates equilibrium states with opposite rotation through $\varrho_J(\varpi)=\Theta\varrho_J(-\varpi)\Theta^{-1}$, from which we find  
\begin{eqnarray}
\rho^{ij}(x,x^{\prime}) & = & {\rm Tr}\varrho_J(\varpi)[\xi^{i}(x),\xi^{j}(x^{\prime})]= \left({\rm Tr}\varrho_J(-\varpi)[\xi^{i}(-x_{0},\boldsymbol{x}),\xi^{j}(-x_{0}^{\prime},\boldsymbol{x}^{\prime})]\right)^{*}= -\rho_{(-\varpi)}^{ij}(-x_{0},\boldsymbol{x};-x_{0}^{\prime},\boldsymbol{x}^{\prime}) ,
\end{eqnarray}
where ${\rm Tr}\Theta^{-1}A\Theta={\rm Tr}A^{*}$. After integrating over $\boldsymbol{p}$ and dropping the $X$ dependence in phase space, we find 
\begin{eqnarray}
{\rm Re}\rho^{ij}(p_{0};\varpi) & = & {\rm Re}\rho^{ij}(p_{0};-\varpi),\quad\quad
{\rm Im}\rho^{ij}(p_{0};\varpi)  =  -{\rm Im}\rho^{ij}(p_{0};-\varpi).\label{eq:Onsager-relation}
\end{eqnarray}
We find the real/symmetric part is an even function of vorticity $2\varpi_{ij}$
while the imaginary/anti-symmetric part is odd. The relation (\ref{eq:Onsager-relation}) is a manifestation of the Onsager reciprocal relation \cite{Onsager:1931kxm}. Because the $\varpi$-dependence is \textit{analytical} order by order, the anti-symmetric part
$\rho^{[ij]}$ is at least of $\mathcal{O}(\varpi)$ and vorticity correction to the conventional
drag-force coefficient $\gamma^{(ij)}$ is of $\mathcal{O}(\varpi^2)$.

A similar analysis of the statistical propagator $G^{F,\mu\nu}(x,x^{\prime})  =  \frac{1}{2}\langle\{\xi^{\mu}(x),\xi^{\nu}(x^{\prime})\}\rangle$ suggests the real (imaginary) part of $G^{F}_{ij}(p,X)$ is (anti-)symmetric with respect to
the $i,j$ indices  and an even (odd) function of $p$, respectively. And time-reversal symmetry further constrains its (anti-)symmetric part to be
an even (odd) function of vorticity $2\varpi_{ij}$, respectively. 

\section{Second-order expansion of the phase-space KMS relation}\label{sec:second_order_KMS}
In studies of spin polarization in relativistic heavy-ion collisions, $\varpi_{\mu\nu}$ is also referred to as the spin chemical potential and can be assumed to be small.
We therefore expand the exact phase-space KMS relation in power series of $\varpi$.
We illustrate this expansion for a vector field, retaining terms through second order in $\varpi$.
Higher-order terms can be derived using the same iterative procedure.
The generalization
to other spin-$j$ fields can be obtained by changing the homogeneous
Lorentz-group spin generator $R_{ab}=(I^{\alpha\beta})_{ab}\varpi_{\alpha\beta}$. 
For vector fields $(I^{\alpha\beta})^{\rho\sigma}=i\eta^{\rho[\alpha}\eta^{\beta]\sigma}$
and for spin-half spinor fields $I^{\alpha\beta}=\frac{1}{2}\sigma^{\alpha\beta}$. For fermions, the statistical signs and thermal factors must also be treated consistently.

Using the Zassenhaus formula up to second order
in $\varpi$,
we factorize $e^{-\varpi_{\rho\sigma}\mathcal{L}^{\rho\sigma}}$ in Eq.~(\ref{eq:New_G}) as
\begin{eqnarray}
e^{-\varpi_{\rho\sigma}\mathcal{L}^{\rho\sigma}} & = & \exp\left(-2\varpi_{\rho\sigma}X^{\rho}p^{\sigma}\right)\exp\left(-\frac{1}{2}\varpi_{\rho\sigma}\partial_{p}^{\rho}\partial_{X}^{\sigma}\right)\exp\left[-\frac{1}{2}\left(\varpi_{\rho\sigma}\varpi^{\rho\sigma}-\varpi_{\rho\sigma}\varpi_{\;\beta}^{\sigma}(X^{\rho}\partial_{X}^{\beta}+p^{\rho}\partial_{p}^{\beta})\right)\right].
\end{eqnarray}
The first factor produces the $X$-dependent correction to the distribution function. We partially resum the $\varpi$ dependence by retaining
$e^{-2\varpi_{\rho\sigma}X^{\rho}p^{\sigma}}$ to all orders,
while expanding the remaining factors through second order in $\varpi$.
For example, the complex translation vector is expanded as
$\tilde{b}_{\mu}=b_{\mu}-i(\varpi b)_{\mu}-\frac{2}{3}(\varpi^{2}b)_{\mu}+...$,
with the shorthand contractions $(\varpi b)_{\mu}=\varpi_{\mu\nu}b^{\nu}$,
$(\varpi^2 b)_{\mu}=\varpi_{\mu\alpha}\varpi^{\alpha\nu}b_{\nu}$,
and $(p\varpi b)=p_\mu\varpi^{\mu\nu}b_{\nu}$.
In the bulk approximation, the $X$ dependence of the Green functions is generated by $\varpi$.
We therefore count derivatives with respect to $X$ as first order in $\varpi$.
The shifted spectral function can then be expanded as
\begin{eqnarray}
\rho_{\mu\nu}\left(p,X-\frac{i}{2}\tilde{b}\right) & = & \rho_{\mu\nu}(p,X)-\frac{i}{2}\tilde{b}\cdot\partial_{X}\rho_{\mu\nu}(p,X)+\frac{1}{2!}\left(-\frac{i}{2}\tilde{b}\cdot\partial_{X}\right)^{2}\rho_{\mu\nu}(p,X)+\mathcal{O}(\partial_X^3).
\end{eqnarray}
To this end, we have 
\begin{eqnarray}
G_{\mu\nu}^{<} & = & G_{\mu\nu}^{<,(0)}+G_{\mu\nu}^{<,(1)}+G_{\mu\nu}^{<,(2)}+\mathcal{O}(\varpi^3) \;,
\end{eqnarray}
where 
\begin{eqnarray}
G_{\mu\nu}^{<,(0)} & = & n(p,X)\rho_{\mu\nu}(p,X) ,\label{eq:G(0)}
\end{eqnarray}
and 
\begin{eqnarray}
G_{\mu\nu}^{<,(1)} & = & 2\left[1+n(p,X)\right]i\varpi_{\mu}^{\;\alpha}G_{\alpha\nu}^{<}(p,X)+\left[1+n(p,X)\right]\left(-i\varpi_{\alpha\beta}p^{\beta}\partial_{p}^{\alpha}+\frac{1}{2}\varpi_{\rho\sigma}\partial_{p}^{\rho}\partial_{X}^{\sigma}+i\varpi_{\rho\sigma}X^{\rho}\partial_{X}^{\sigma}\right)G_{\mu\nu}^{<}(p,X)\nonumber \\
 &  & +n(p,X)\left(\frac{-i}{2}b\cdot\partial_{X}+i(p\varpi b)\right)\left(\rho_{\mu\nu}(p,X)+G_{\mu\nu}^{<}(p,X)\right) ,
\end{eqnarray}
and 
\begin{eqnarray}
 G_{\mu\nu}^{<,(2)}(p,X)
 & = & \left[1+n(p,X)\right]\left\{ \frac{1}{2}\varpi_{\rho\sigma}\varpi_{\alpha\beta}p^{\sigma}p^{\beta}\partial_{p}^{\rho}\partial_{p}^{\alpha}G_{\mu\nu}^{<}(p,X)-2\varpi_{\mu}^{\;\alpha}\left(\varpi_{\rho\sigma}p^{\sigma}\partial_{p}^{\rho}\right)G_{\alpha\nu}^{<}(p,X)+2(\varpi^{2})_{\mu}^{\;\alpha}G_{\alpha\nu}^{<}(p,X)\right\} \nonumber \\
 &  & +n(p,X)\left[\frac{-1}{2}(\varpi b)\cdot\partial_{X}+\frac{-1}{8}(b\cdot\partial_{X})^{2}+\frac{1}{2}(p\varpi b)b\cdot\partial_{X}-\frac{1}{2}(p\varpi b)^{2}+\frac{2}{3}(p\varpi^{2}b)+\frac{1}{2}\varpi_{\rho\sigma}\varpi^{\rho\sigma}\right]\nonumber\\
 & &\quad\times\left(\rho_{\mu\nu}(p,X)+G_{\mu\nu}^{<}(p,X)\right) .
\end{eqnarray}
The Bose-Einstein distribution with vorticity correction that automatically
emerges from the phase-space orbital angular momentum operator is
defined as
\begin{eqnarray}
n(p,X) & = & \frac{1}{e^{\beta(X)\cdot p-\alpha}-1},
\end{eqnarray}
where $\alpha=\mu/T$ as defined in the main text and $\beta_{\mu}(X)=b_{\mu}+2\varpi_{\mu\nu}X^{\nu}$ 
is the position-dependent inverse-temperature four-vector.
Equation~(\ref{eq:G(0)}) has the conventional KMS form,
but with the position-dependent distribution $n(p,X)$.
Higher-order corrections are obtained by substituting lower-order Wightman functions
into the implicit expressions above.
For example, we begin by substituting Eq.~(\ref{eq:G(0)}) into the first two terms of $G_{\mu\nu}^{<,(1)}$.
The term $b\cdot\partial_{X}G_{\mu\nu}^{<}$ still involves the full Wightman function and is treated iteratively. Writing the remaining terms as $\mathcal{F}_{\mu\nu}(p,X)$, we obtain
\begin{eqnarray}
G_{\mu\nu}^{<}(p,X)  =  \mathcal{F}_{\mu\nu}(p,X)-\frac{i}{2}n(p,X)b\cdot\partial_{X}G_{\mu\nu}^{<}(p,X)+... = \mathcal{F}_{\mu\nu}(p,X)-\frac{i}{2}n(p,X)b\cdot\partial_{X}\mathcal{F}_{\mu\nu}(p,X)+...
\end{eqnarray}
The hermiticity properties discussed in Sec.~\ref{sec:symmetry-properties}
also constrain the symmetric and antisymmetric parts of the Green functions.
Combining these considerations with the iterative expansion gives,
through $\mathcal{O}(\varpi^{2})$,
\begin{eqnarray}
 G_{\mu\nu}^{<}(p,X)
 & = & \left[n-\frac{1}{2}n_{(2)}\varpi^{\alpha\beta}\varpi_{\alpha\beta}-\frac{1}{3}n_{(3)}(p_{\sigma}\varpi^{\sigma\alpha}\varpi_{\alpha\beta}b^{\beta})\right]\rho_{\mu\nu}+in_{(1)}\{\varpi,\rho\}_{\mu\nu}-n_{(2)}\{\varpi^{2},\rho\}_{\mu\nu}\nonumber\\
 &  & +\frac{1}{2}n_{(2)}(\varpi^{2}p)_{\lambda}\partial_{p}^{\lambda}\rho_{\mu\nu}-\frac{1}{2}nn_{(2)}(b\varpi p)\varpi_{\rho\sigma}p^{\sigma}\partial_{p}^{\rho}\rho_{\mu\nu}+\frac{1}{2}n_{(1)}\varpi_{\alpha\beta}\partial_{p}^{\alpha}\partial_{X}^{\beta}\rho_{\mu\nu}+\mathcal{O}(\varpi^3),\label{eq:VF-KMS}
\end{eqnarray}
where the tensor contraction is defined as before, $(\varpi\rho)_{\alpha\beta}=\varpi_{\alpha\gamma}\rho_{\;\beta}^{\gamma}$.
We denote the thermal functions $n_{(1)}=n(1+n)$, $n_{(2)}=n_{(1)}\left(1+2n\right)$
and $n_{(3)}=n_{(1)}\left(1+6n_{(1)}\right)$.  Interestingly, the hermiticity of $G^<_{\mu\nu}$ gives the following equation of motion for rotational spectral function 
\begin{equation}
\mathfrak{L}_{b}\rho_{\mu\nu}=2\varpi_{\rho\beta}(p^{\rho}\partial_{p}^{\beta}+X^{\rho}\partial_{X}^{\beta})\rho_{\mu\nu}+\mathcal{O}(\varpi^{3}),
\end{equation}
where $\mathfrak{L}_b=b^{\mu}\partial_{X,\mu}$ is the Lie derivative along the timelike vector $b^{\mu}$. This means that the change in the spectral function along the thermal-bath velocity in a rotating system
corresponds to its rotational dilation in phase space.
Equation~(\ref{eq:VF-KMS}) does not require a specific interaction model or a weak-coupling expansion.
It is therefore nonperturbative in the interaction strength
and can be applied to strongly coupled systems such as the quark-gluon plasma.
The result retains both the rotation and boost components of $\varpi_{\mu\nu}$,
allowing acceleration-induced corrections to be studied in interacting systems.
Its connection to the Unruh effect~\citep{Crispino:2007eb} requires a separate analysis.
For pure rotation, where $b_{\mu}\varpi^{\mu\nu}=0$ and $\tilde{b}=b$, Eq.~(\ref{eq:VF-KMS}) can be simplified considerably.


\section{Transport properties of heavy particle in rotating medium}
\label{sec:Langevin_equation}

The Einstein relation constrains momentum diffusion and drag in the low-frequency transport limit.
In this section, we discuss its generalization to a rotating medium and  the contributions of nonanalytic terms
in the spectral and statistical functions.
Consider the heavy particle in a bath,  with Hamiltonian $\hat{H}=\hat{H}_{\text{S}}+\hat{H}_{\text{E}}+\hat{H}_{\text{int}}$.
Here $\hat{H}_{\text{S}}=\frac{\hat{\boldsymbol{p}}^{2}}{2M}+V(\hat{\boldsymbol{x}})$
where $V(\hat{\boldsymbol{x}})$ is the heavy-particle potential  and $\hat{H}_{\text{E}}$ is the Hamiltonian for the environment. The
interaction between the heavy particle and the environment is governed
by a random force kick $\hat{H}_{\text{int}}=-\hat{\boldsymbol{x}}\cdot\hat{\boldsymbol{\xi}}$, where $\hat{\boldsymbol{\xi}}$ is the bath-force operator.
We neglect the spatial dependence of the bath force
in the local near-axis approximation for a slowly rotating medium.
Then in the Heisenberg
picture, with the linear response of the noise operator \citep{Kapusta:2006pm} and the Markovian approximation in the kick kernel, we derive the time-honored quantum Langevin equation \citep{gardiner2000quantum},
\begin{eqnarray}
\frac{{\rm d}}{{\rm d}t}p^{i}(t) & = & F_{\text{con}}^{i}(\boldsymbol{x},t)+\xi^{i}(t)-M\eta_{{\rm D}}^{ij}\frac{{\rm d}}{{\rm d}t}x^{j}(t) ,\label{eq:Langevin-eq}
\end{eqnarray}
where $\xi^{i}(t)$ is the noise function.
We define the retarded noise correlator $G_{{\rm R},\xi}^{ij}(t,t^{\prime})=i\theta(t-t^{\prime})\langle[\hat{\xi}_{{\rm I}}^{i}(t),\hat{\xi}_{{\rm I}}^{j}(t^{\prime})]\rangle_{\text{E}}$ where the subscript ${\rm I}$ denotes the interaction picture, $\langle O\rangle_{\text{E}}=\text{Tr}\hat\varrho_{\text{E}} \hat{O}$. Here $\hat\varrho_{\text{E}}$ is the environment density operator.
The effective \textit{conservative} external force is 
\begin{eqnarray}
F_{\text{con}}^{i}(\boldsymbol{x},t) & = & -\boldsymbol{\nabla}_{\boldsymbol{x}}^{i}V(\boldsymbol{x})+x^{j}(t)\lim_{E\to0}G_{{\rm R},\xi}^{ij}(E) ,\label{eq:Conservative-force}
\end{eqnarray}
where the subscript $\xi$ indicates that $G_{R, \xi}^{ij}$ is the retarded correlator of the force $\xi$.
Later, we use the subscripts $EE$ and $AA$ for correlators of the electric field $\boldsymbol{E}$
and the gauge field $A$, respectively. 
More importantly, the drag-force coefficient is
\begin{eqnarray}
\eta_{{\rm D}}^{ij} & = & {\rm Re}\frac{1}{M}\int_{-\infty}^{+\infty}{\rm d}t^{\prime}G_{{\rm R},\xi}^{ij}(t,t^{\prime})(t-t^{\prime})=\frac{1}{M}{\rm Im}\lim_{E\to0}\partial_{E}G_{{\rm R},\xi}^{ij}(E),\label{eq:Drag-force-coefficient}
\end{eqnarray}
where in the second step we use stationarity
and neglect the explicit spatial dependence near the rotation axis.

In a rotational medium with a global vorticity $\boldsymbol{\omega}_i=2\epsilon_{ijk}T\varpi_{jk}$ as defined in the main text,
the form of the momentum-diffusion Langevin equation (\ref{eq:Langevin-eq})
is not altered. But the conservative potential is modified by the vorticity potential $\delta V(\boldsymbol{x},\boldsymbol{p})=-\boldsymbol{\omega}\cdot\langle\hat{\boldsymbol{x}}(t)\times\hat{\boldsymbol{p}}(t)\rangle-\boldsymbol{\omega}\cdot\boldsymbol{S}$.
It brings two effects. First, the equation of motion obtained from the Ehrenfest theorem \citep{Sakurai:2011zz}
is changed by the orbital-vorticity coupling $\partial_t x^{j}(t)  =  p^{j}(t)/M-\epsilon^{jmn}\omega^{m}x^{n}(t)+...$.
Second, the effective conservative external force in Eq.~(\ref{eq:Conservative-force})
receives a correction $-\boldsymbol{\omega}\times\boldsymbol{p}$.
These two effects can actually be discarded in our discussions of
Einstein relation. Since we are interested in the near-axis region
of the slowly rotating system, $\boldsymbol{\omega}\times\boldsymbol{x}(t)$
is a small effect and can be neglected, so the Langevin equation
remains same. On the other hand, the rotation potential is
a conservative source and does not influence the dissipative dynamics, i.e.,
the diffusion coefficient $\eta_{{\rm D}}^{ij}$.

As a result, the rotation effects on the dissipative Langevin dynamics lie in the environment correlations.
Mapping of Eq.~(\ref{eq:Langevin-eq})
to the Fokker-Planck equation defines the diffusion coefficient $\kappa^{ij}$
in terms of the noise correlator $\langle\xi^{i}(t)\xi^{j}(t^{\prime})\rangle=\kappa^{ij}\delta(t-t^{\prime})$ with $\kappa^{ij}$ symmetric in its $i,j$ indices. Equivalently,
\begin{eqnarray}
\kappa^{ij} & = & \frac{1}{2}(\kappa^{ij}+\kappa^{ji})=\frac{1}{2}\lim_{E\to0}G_{\xi}^{F,(ij)}(E). \label{eq:Diffusion-coefficient}
\end{eqnarray}
The Einstein relation relates
different transport coefficients (\ref{eq:Drag-force-coefficient})
and (\ref{eq:Diffusion-coefficient}) through the FDT and time-translation symmetry. In a rotating system, both the FDT and the Einstein relation are modified by vorticity. To see this, we start from the transport limit $E\to0$ of rotating KMS relation.
We define $\rho_{\xi}^{ij}(E)=\int\frac{{\rm d}^{3}\boldsymbol{p}}{(2\pi)^{3}}\rho_{\xi}^{ij}(E,\boldsymbol{p})$.
With Eq.~(\ref{eq:VF-KMS}) the KMS relation at $\mathcal{O}(\varpi^{2})$ gives
\begin{eqnarray}
G_{\xi}^{F,ij}(E) & = & \rho^{ij}_{\xi}(E)\left[n(E)+\frac{1}{2}\right]-i\left[\varpi^{im}\rho^{mj}_{\xi}(E)+\rho^{im}_{\xi}(E)\varpi^{mj}\right]n_{(1)}(E)\nonumber \\
 &  & \quad-\left[\varpi^{im}\varpi^{mn}\rho^{nj}_{\xi}(E)+\rho^{im}_{\xi}(E)\varpi^{mn}\varpi^{nj}\right]n_{(2)}(E)+\mathcal{O}(\varpi^3) ,\label{eq:statistical-function-1}
\end{eqnarray}
where we have used integration by parts and dropped the total derivative terms
during $\boldsymbol{p}$ integration. 
In general, the IR region of the $\boldsymbol{p}$-integrated spectral function $\rho_{\xi}^{ij}(E)$ can be non-analytic. A simple Taylor expansion does not exist because the fixed-$E$ expansion of $\rho^{ij}_{\xi}(E,\boldsymbol{p})$ can develop an infrared divergence \citep{Burnier:2010rp} and the $\boldsymbol{p}$-integration and $E$ expansion are not interchangeable. 
We now write the $E\to0$ behavior of $\rho_{\xi}^{ij}(E)$
as the sum of its Taylor-expanded and non-analytic parts,
\begin{eqnarray}
\rho_{\xi}^{ij}(E) & = & i\gamma_{0}^{ij}+2E\gamma_{1}^{ij}+iE^{2}\gamma_{2}^{ij}+E^{3}\gamma_{3}^{ij}+\gamma_{\text{na}}^{ij}(E)+\mathcal{O}(E^{4}),
\end{eqnarray}
where all the coefficients $\gamma_{n}^{ij}$ are real; $\gamma_{2n}^{ij}$
are anti-symmetric whereas $\gamma_{2n+1}^{ij}$ are symmetric
for $n=0,1,\ldots$. Inserting this expansion into Eq.~(\ref{eq:statistical-function-1})
and taking the $E\to0$ limit, we express $\kappa^{ij}$ in terms of spectral function modes $\gamma_{n}^{ij}$,
\begin{eqnarray}
    \kappa^{ij} &=& \frac{1}{2}\lim_{E\to0}G_{\xi}^{F,(ij)}(E) \nonumber \\ 
    &=&2T\gamma_{1}^{ij}+\gamma_{2}^{m(i}\varpi^{j)m}T^{2}+\varpi^{m(i}\gamma_{3}^{j)n}\varpi^{mn}2T^{3}\nonumber\\
 &  & +\left[\varpi^{im}(\gamma_{0}^{mj}-4T\varpi^{mn}\gamma_{1}^{nj})+(\gamma_{0}^{im}+4T\gamma_{1}^{in}\varpi^{mn})\varpi^{mj}\right]\frac{T^{2}}{E^{2}}+\text{non-analytic part}.
\end{eqnarray}
By definition $M\eta_{{\rm D}}^{(ij)}=2\gamma_{1}^{ij}$.
The Onsager relation (\ref{eq:Onsager-relation}),
together with isotropy in the nonrotating limit, implies
$2\gamma_{1}^{ij}=2M\eta_{{\rm D}}\delta^{ij}+\mathcal{O}(\varpi^{2})$.
Using these relations and the finiteness of $\kappa^{ij}$,
we obtain the following Einstein relations up to $\mathcal{O}(\varpi^{2})$,
\begin{eqnarray}
\kappa^{ij} & = & TM\eta_{{\rm D}}^{(ij)}+\gamma_{2}^{m(i}\varpi^{j)m}T^{2}-2T^{3}\varpi^{nm}\varpi^{m(i}\gamma_{3}^{j)n} ,\\
\gamma_{0}^{mj} & = & 4TM\eta_{{\rm D}}\varpi^{mj}.\label{eq:Constraint-2}
\end{eqnarray}
This means that $\kappa^{ij}$ is modified by vorticity and that there is a new Einstein-type
constraint between $\gamma_{0}$ and $\gamma_{1}$. We remark that $\gamma_{\text{na}}^{ij}(E)$ does not make contributions to $\kappa^{ij}$. 
For the non-analytic part, consider terms proportional to $E^{\lambda}$ with noninteger $\lambda$.
Then in the IR region, $E^{\lambda}/E^{n}$ vanishes for $\lambda>n$,
while terms with $\lambda<n$ give an IR singularity.
This singularity must cancel against a contribution with the same IR scaling
from the corresponding symmetric or antisymmetric sector,
yielding constraints similar to Eq.~(\ref{eq:Constraint-2}).


\section{An example of spectral zero modes of a rotating QED plasma}\label{sec:zero_modes}
 
We calculate the transport coefficients for a heavy fermion propagating
in a hot quantum electrodynamics (QED) plasma with rotation. 
In the nonrelativistic limit considered above,
the Lorentz force reduces to $\boldsymbol{\xi}(t)=e\boldsymbol{E}(t)$.
The force spectral function is therefore determined by the electric-field correlator
\begin{eqnarray}
\rho_{\xi}^{ij}(x,y) & = & e^{2}\langle[E^{i}(x),E^{j}(y)]\rangle\equiv e^{2}\rho_{EE}^{ij}(x,y).\label{eq:noise-spectral-function}
\end{eqnarray}
The electric-field spectral function is related to the retarded and advanced correlators by
$i\rho_{EE}^{ij}=G_{EE}^{{\rm R},ij}-G_{EE}^{{\rm A},ij}$,
where $G_{EE}^{{\rm R/A},ij}(x,y)=\pm i\theta[\pm(x_{0}-y_{0})]\rho_{EE}^{ij}(x,y)$.
In thermal perturbation theory, all the $X$-dependence of Wigner
functions is built from their free counterparts. We find from Eq.~(\ref{eq:VF-KMS}) that the
spacetime-dependent distribution functions e.g. $n(p,X)$ depend only
on $X^{i}$ in the pure rotation case. For convenience, we work with the temporal gauge $A_{0}=0$ and the electric spectral function
Eq.~(\ref{eq:noise-spectral-function})
can be solely expressed solely in terms of the $ij$ gauge-field spectral function
\begin{eqnarray}
\rho_{EE}^{ij}(q,X) & = & q_{0}^{2}\rho_{AA}^{ij}(q,X).
\end{eqnarray}

We calculate the rotational correction to the spectral function
at leading-logarithmic order in a high-temperature weakly-coupled plasma. 
The equation of motion for the retarded Wigner function
is the Keldysh equation \citep{Fang:2025pzy}, and 
we treat $\varpi$ perturbatively. We denote the Dyson-resummed
propagator $G_{{\rm R/A},0}$ as the ``vorticity-free'' part and $\Sigma_{\rm{R/A},\varpi}$ as the ``vorticity-corrected'' self energy.  Up to $\mathcal{O}(\varpi)$, 
\begin{eqnarray}
&  G_{{\rm R/A}}^{\mu\nu}(q,X)& \approx   G_{{\rm R/A},0}^{\mu\nu}(q)-G_{{\rm R/A},0}^{\mu\beta}(q)\Sigma_{\beta\alpha,{\rm R/A},\varpi}(q,X)G_{{\rm R/A},0}^{\alpha\nu}(q)\nonumber \\
 &  & +\frac{i}{2}\left[G_{{\rm R/A},0}^{\mu\beta}(q)\partial_{X}^{\lambda}\Sigma_{,\beta\alpha{\rm R/A},\varpi}(q,X)\partial_{q,\lambda}G_{{\rm R/A},0}^{\alpha\nu}(q)-\partial_{q,\lambda}G_{{\rm R/A},0}^{\mu\beta}(q)\partial_{X}^{\lambda}\Sigma_{\beta\alpha,{\rm R/A},\varpi}(q,X)G_{{\rm R/A},0}^{\alpha\nu}(q)\right], \nonumber \\ \label{eq:Keldysh-eq}
\end{eqnarray}
where every $\partial_{X}$ generates a term of $\mathcal{O}(\varpi)$
since $\partial_{X}n\sim(\varpi p)n_{(1)}$. The HTL retarded/advanced photon self-energy in the rotating medium,
including terms through first order in $\varpi$, is
\begin{eqnarray}
\Sigma_{{\rm R/A}}^{\beta\alpha}(q,X) & = & e^{2}\int\frac{{\rm d}^{4}p}{(2\pi)^{4}}{\rm Tr}\left[\gamma^{\alpha}S_{{\rm A/R}}(p-q)\gamma^{\beta}S_{F}(p,X)+\gamma^{\beta}S_{{\rm R/A}}(q+p)\gamma^{\alpha}S_{F}(p,X)\right]\nonumber\\
& \simeq & e^{2}\int\frac{{\rm d}\Omega_{v}}{4\pi}\left[\frac{q^{\alpha}v^{\beta}+q^{\beta}v^{\alpha}}{v\cdot q\pm i\eta}-\frac{v^{\beta}v^{\alpha}q^{2}}{(v\cdot q\pm i\eta)^{2}}-\eta^{\alpha\beta}\right]\left(\frac{T^{2}}{6}-2\varpi^{\rho\sigma}v_{\rho}X_{\sigma}\frac{T^{3}}{3}\right)\nonumber \\
 &  & +\frac{e^{2}T^{3}}{3}i(\varpi^{\mu\beta}\eta^{\alpha\delta}-\varpi^{\mu\alpha}\eta^{\beta\delta})\int\frac{{\rm d}\Omega_{v}}{4\pi}\frac{v_{\delta}v_{\mu}}{v\cdot q\pm i\eta}+\mathcal{O}(\varpi^{2}).
\end{eqnarray}
where $v^{\mu}=u^{\mu}+\hat{p}_{\perp}^{\mu}$ and $\hat{p}_{\perp}^{\mu}=p_{\perp}^{\mu}/E_{p}$
with $p_{\perp}^{\mu}=(\eta^{\mu\nu}-u^{\mu}u^{\nu})p_{\nu}$ and $u^{\mu}=b^{\mu}/\sqrt{b^{2}}$. 
We take the light fermions in the plasma to be massless and neglect their chemical potential.
Here $S_{{\rm R/A}}$ and $S_F$ denote their free retarded/advanced and statistical propagators,
respectively.
The photonic momentum $q\sim\mathcal{O}(eT)$ is much smaller than the fermion-loop momentum $p\sim\mathcal{O}(T)$.
The $\mathcal{O}(\varpi^{0})$ self-energy recovers the rotationless photonic HTL propagator \citep{Kapusta:2006pm} and its $\mathcal{O}(\varpi)$ part satisfies
the phase-space Ward identity 
\begin{equation}
q_{\alpha}\Sigma_{{\rm R/A}}^{\beta\alpha}(q,X)  =  \frac{i}{2}\partial_{X,\alpha}\Sigma_{{\rm R/A}}^{\beta\alpha}(q,X).
\end{equation}

In the transport region of the high-temperature limit, where the Landau-cut
contribution dominates the spectral function, we find that the leading
$q_{0}$ behavior of $\rho_{\xi}$ is
\begin{eqnarray}
\rho_{\xi}^{ij}(q_0) & = & e^2q_{0}^{2}\rho_{AA,0}^{ij}(q_{0})=\frac{1}{3}e^2q_{0}\delta^{ij}\int\frac{{\rm d}^{3}\boldsymbol{q}}{(2\pi)^{3}}\frac{|\boldsymbol{q}|2\pi m_{{\rm D}}^{2}}{\left(\boldsymbol{q}^{2}+2m_{{\rm D}}^{2}\right)^{2}}+ 2ie^2T\varpi_{ij}\int\frac{{\rm d}^{3}\boldsymbol{q}}{(2\pi)^{3}}\frac{m_{{\rm D}}^{2}|\boldsymbol{q}|\frac{2}{3}\pi}{\left(\boldsymbol{q}^{2}+2m_{{\rm D}}^{2}\right)^{2}}+o(q_0)\nonumber\\
 & = &i\varpi_{ij}\frac{T^{3}}{9\pi}e^{4}\log e^{-1}+q_{0}\delta^{ij}\frac{T^{2}}{18\pi}e^{4}\log e^{-1}+o(q_0) ,
\end{eqnarray}
where the Debye mass is $m_{{\rm D}}^{2}=e^{2}T^{2}/6$. We find the IR behavior of the spectral function indeed satisfies the general non-perturbative
constraint (\ref{eq:zero_modes}) $\gamma_{0}^{ij}=4MT\eta_{{\rm D}}\varpi^{ij}$ with $\gamma_{0}^{ij}=\varpi^{ij}\frac{T^{3}}{9\pi}e^{4}\log e^{-1}$
and $\eta_{{\rm D}}=\frac{T^{2}}{36\pi M}e^{4}\log e^{-1}$. 

\section{Two examples of orbital polarization in a rotating medium}\label{sec:orbtail_polarization}

We derive the evolution of mean orbital angular momentum for heavy
quarkonium in the rotating medium based on an open-quantum-system approach
\citep{breuer2002theory,Akamatsu:2020ypb,Yao:2021lus}.
For simplicity we work in the high-temperature limit such that the
binding energy $E$ of heavy quarkonium is small compared to the medium
temperature $T$ \citep{Brambilla:2017zei}.
For phenomenological illustration, we expand to leading order in $E/T$.
Vorticity corrections to the heavy-quark potentials can be neglected
when they are of $\mathcal{O}(T\varpi)$ and are much smaller than
the typical binding energy $E$ \citep{Brambilla:2004jw}.
We assume this hierarchy in the following. We keep the subscript $J$ explicit in this section to avoid confusion.
For example, we use $\varrho(t)$ and $\varrho_{J}(t)$ to represent the density operators
in the nonrotating and rotating media, respectively.
The collision terms in the quantum master equations of Ref.~\citep{Brambilla:2017zei} for the color-singlet/octet density operators $\varrho_{J,s/o}$
receive corrections proportional to $\gamma_0^{ij}$, 
\begin{eqnarray}
\delta C_{s}(t)  & = & -\frac{3i}{2}\gamma_{0}^{ij}\boldsymbol{r}^{i}\frac{\varrho_{J,o}(t)}{N_{c}^{2}-1}\boldsymbol{r}^{j},\quad \delta C_{o}(t) = -\frac{3i}{2}\gamma_{0}^{ij}\boldsymbol{r}^{i}\left(\frac{N_{c}^{2}-4}{2}\frac{\varrho_{J,o}(t)}{N_{c}^{2}-1}+\varrho_{J,s}(t)\right)\boldsymbol{r}^{j}.
\end{eqnarray}
Here the anti-symmetric transport coefficient $\gamma_0^{ij}$ is defined in terms of an adjoint chromoelectric-field correlator with implicit Wilson-line insertion,
\begin{eqnarray}
i\gamma_{0}^{ij} & = & \frac{g^{2}T_{{\rm F}}}{3N_{c}}\int_{t_{0}}^{t}{\rm d}t^{\prime}\left[\langle[\boldsymbol{E}^{a,i}(t),\boldsymbol{E}^{a,j}(t^{\prime})]\rangle_{J,E}-\langle[\boldsymbol{E}^{a,j}(t),\boldsymbol{E}^{a,i}(t^{\prime})]\rangle_{J,E}\right],
\end{eqnarray}
where $T_F$ is the trace normalization of the fundamental generators,
and the expectation value $\langle \cdots \rangle_{J,E}$ used above is evaluated with
the environment density operator $\varrho_{J,\text{E}}$ in the presence of thermal vorticity.
Such corrections only make subleading contribution to the evolution of the color-singlet/octet density operators $\varrho_{J,s/o}$. Their contributions can be isolated in angular-momentum-resolved dynamics.
Chromomagnetic dipole interactions that flip the heavy-quark spin
are suppressed in the heavy-mass expansion.
We therefore focus on orbital angular momentum at the order considered here.

In an isotropic medium, the orbital angular momentum (OAM) $L^{k}$ is distributed uniformly and thus ${\rm Tr}\varrho_{s/o}L^{k}r^{2}=0$
as there is no preferred direction in the environment. This quantity does not change with time.
In a rotating medium ${\rm Tr}\varrho_{J, s/o}L^{k}r^{2}\neq0$, and we derive an evolution equation for the OAM of a color-singlet heavy-quark pair from the evolution of $\varrho_{J,s}(t)$,  
\begin{eqnarray}
\frac{{\rm d}}{{\rm d}t}\langle L^{k}\rangle_{J, s} & = & \kappa{\rm Tr}\left(\frac{\varrho_{J, o}(t)}{N_{c}^{2}-1}-\varrho_{J, s}(t)\right)L^{k}r^{2}-\frac{3}{2}\epsilon^{jkm}\gamma_{0}^{ij}{\rm Tr}\frac{\varrho_{J, o}(t)}{N_{c}^{2}-1}r^{m}r^{i} ,\label{eq:OAM-evolution-1}
\end{eqnarray}
where $\langle O\rangle_{J, s/o}=\text{Tr}\varrho_{J, s/o}O$ and $\kappa=\delta_{ij}\kappa^{ij}$ is the heavy-quarkonium diffusion coefficient. Up to first order in vorticity, the first term of Eq.~(\ref{eq:OAM-evolution-1}) can be expressed in terms of quantities averaged with non-rotating density operators $\varrho_{s/o}$. To conclude,
\begin{eqnarray}
\frac{{\rm d}}{{\rm d}t}\langle L^{k}(t)\rangle_{J, s} & = & \gamma_{0}^{ij}\left[C^{ijk}(t)-\epsilon^{ijk}\frac{\langle\boldsymbol{r}^{2}(t)\rangle_{o}}{2(N_{c}^{2}-1)}\right]. \label{eq:orbtial_2}
\end{eqnarray} 
where $\langle\boldsymbol{r}^{2}(t)\rangle_{o}\equiv \text{Tr} \varrho_o(t) \boldsymbol{r}^2$ and we introduce
\begin{eqnarray}
C^{ijk}(t) & = & \frac{3i\kappa}{2(N_{c}^{2}-1)}\int_{t_{0}}^{t}{\rm d}t^{\prime}{\rm Tr}\left(\varrho_{o}(t^{\prime})\boldsymbol{r}^{j}L^{k}\boldsymbol{r}_{s}^{2}(t,t^{\prime})\boldsymbol{r}^{i}-\left[\frac{N_{c}^{2}-4}{2(N_{c}^{2}-1)}\varrho_{o}(t^{\prime})+\varrho_{s}(t^{\prime})\right]\boldsymbol{r}^{j}L^{k}\boldsymbol{r}_{o}^{2}(t,t^{\prime})\boldsymbol{r}^{i}\right) ,
\end{eqnarray}
and $\boldsymbol{r}_{s/o}^{2}(t,t^{\prime})=e^{ih_{s/o}^{{\rm eff}}(t-t^{\prime})}\boldsymbol{r}^{2}e^{-ih_{s}^{{\rm eff}}(t-t^{\prime})}$. 
Here $h_{s/o}^{{\rm eff}}=h_{s/o}+\Delta h_{s/o}^{{\rm LS}}$,
where $\Delta h_{s/o}^{{\rm LS}}$ denotes the Lamb shift
of the color-singlet/octet Hamiltonian.
In Eq.~(\ref{eq:orbital_polarization}) of the main text, we omit the subscripts $s$ and $o$
used in Eq.~(\ref{eq:orbtial_2}) to avoid notational clutter. 
The rotating KMS relation (\ref{eq:VF-KMS}) reduces the number of independent transport
coefficients in spin transport, so the evolution of $L^{k}$ depends only on the conventional transport coefficient $\kappa$. A nonzero
$\langle L^{k}\rangle$ indicates an interplay between fluid
vorticity and the strongly-coupled QCD force.
It provides information about both the QCD interaction of the medium and the collective spin hydrodynamical  modes. 


We now explain the physical meaning of $\langle L^{k}\rangle_{J, s}$.
Take an $l=1$ heavy quarkonium state, for example. The density operator can
be expanded as $\varrho_{J, l=1}=\sum_{m_{1},m_{2}}n_{m_{1}m_{2}}|l=1,m_{1}\rangle_J \langle l=1,m_{2}|_J$
so $\langle L^{z}\rangle_J={\rm Tr}\varrho_{J,l=1}L^{z}=n_{1}-n_{-1}$.  
Here $n_m$ denotes the population of the orbital magnetic sublevel $m$.
Thus, $\langle L^{z}\rangle_J$ is the difference between the occupation numbers
of the $m=1$ and $m=-1$ orbital sublevels of $P$-wave quarkonium,
which characterizes the \textit{polarization of the orbital angular momentum}.
In the radiative decay of spin-singlet quarkonium,
the orbital polarization of the parent is reflected
in the helicity-dependent photon angular distribution.
In the
process $h_{c}\to\eta_{c}+\gamma$, the photon distribution with helicity
$\lambda$ is given by \citep{Kadeer:2005aq}
\begin{eqnarray}
\frac{{\rm d}\Gamma_{h_{c}\to\eta_{c}+\gamma}^{(\lambda)}}{{\rm d}\cos\theta} & \sim & |H_{\lambda}|^{2}\left[n_{1}\frac{(1+\lambda\cos\theta)^{2}}{4}+n_{0}\frac{\sin^{2}\theta}{2}+n_{-1}\frac{(1-\lambda\cos\theta)^{2}}{4}\right],
\end{eqnarray}
Here $\lambda=\pm1$, and $\theta$ is the angle between the photon momentum
and the $z$ axis, chosen along the global vorticity direction as in the main text.
The squared moduli of the helicity amplitudes satisfy
$|H_{+1}|^2=|H_{-1}|^2\equiv|H|^2$ in this parity-conserving electromagnetic decay.
The difference between the two helicity-resolved angular distributions
is therefore proportional to $\langle L^{z}\rangle_{J,s} \cos\theta$, i.e.
$\text{d} \left(\Gamma_{h_{c}\to\eta_{c}+\gamma}^{(\lambda=1)}-\Gamma_{h_{c}\to\eta_{c}+\gamma}^{(\lambda=-1)}\right)/\text{d}\cos\theta  \sim |H|^{2}\left(n_{1}-n_{-1}\right)\cos\theta$.

Photon reconstruction is generally difficult in experiments, but
a rare decay process such as $\chi_{c1}\to\Lambda\bar{\Lambda}$ \citep{BESIII:2025gof} provides
another promising channel. From the Wigner-Eckart theorem, the OAM
of $\chi_{c1}$ is proportional to its total angular momentum according to $\langle L^{z}\rangle_J=\frac{J(J+1)+L(L+1)-S(S+1)}{2J(J+1)}\langle J^{z}\rangle_J$.
For $\chi_{c1}(^{3}P_{1})$ we therefore have $\langle L^{z}\rangle_J=\frac{1}{2}\langle J^{z}\rangle_J$.
In $\chi_{c1}\to\Lambda\bar{\Lambda}$,
charge-conjugation and parity conservation give $\mathcal{P}=1$ and $\mathcal{C}=1$ for the final state.
Using $\mathcal{P}_{{\rm fin}}=(-1)^{L_{{\rm f}}+1}$
and $\mathcal{C}_{{\rm fin}}=(-1)^{L_{{\rm f}}+S_{{\rm f}}}$,
together with total-angular-momentum conservation,
we obtain $J_{{\rm f}}=1$, $L_{{\rm f}}=1$, and $S_{{\rm f}}=1$.
The $\Lambda\bar{\Lambda}$ pair is therefore in a $^{3}P_{1}$ state and forms
a spin triplet with parallel spins. 
For normalized spin density matrices, the Wigner-Eckart theorem gives
the decay-angle-averaged polarization-transfer relation
\begin{eqnarray}
P_{\Lambda}^{z}
& = &
2\langle s_{\Lambda}^{z}\rangle_J=\langle L_{\chi_{c1}}^{z}\rangle_J,
\end{eqnarray}
where all spin projections refer to a common quantization axis in the parent rest frame.

\section{An example of detailed balance with rotation}
\label{sec:detailed_balance}

The rotational detailed balance relation in Eq.~(\ref{eq:Modified-detailed-balance}) of the main text does not rely on
interaction details of the bath.
In this section, we consider a common setup in atomic physics, atoms
in a QED cavity with electric-dipole interactions $\hat{H}_{\text{int}}=-\hat{\boldsymbol{d}}\cdot \hat{\boldsymbol{E}}$. Here $\hat{\boldsymbol{d}}$ is the electric-dipole operator and $\boldsymbol{E}$ is the electric field. The cavity field is assumed to be in rotating thermal equilibrium.
As in the main text, its local electric-field correlator is isotropic in the absence of rotation.
We show the modified detailed balance is indeed satisfied and is consistent with the result derived from the channel-resolved rotating KMS relation.
We do not specify the density of states of the cavity and assume the
energy gaps of the atoms are much larger than the atom-cavity coupling.
Using the standard open-quantum-system framework with Born-Markovian
 and rotating wave approximation, we obtain the time
evolution of the population $N_{n}(t)=\langle n|\hat{\varrho}_{\text{S},J}(t)|n\rangle$,   where $\varrho_{\text{S}(J)}$ is is the density matrix of the atomic system without/with rotation and 
the $|n\rangle$ is an atomic energy eigenstate, from the Lindblad equation \citep{breuer2002theory},
\begin{eqnarray}
\frac{{\rm d}}{{\rm d}t}N_{n}(t) & = & \sum_{m}\langle n|\hat{d}^{\dagger,j}|m\rangle\langle m|\hat{d}^{i}|n\rangle\left[G_{EE}^{>,ij}(E_{m}-E_{n})N_{m}(t)-G_{EE}^{>,ji}(E_{n}-E_{m})N_{n}(t)\right]\nonumber \\
 & \equiv & \sum_{m}\left[\Gamma_{n\to m}(\Delta E)N_{m}(t)-\Gamma_{m\to n}(\Delta E)N_{n}(t)\right] ,
\end{eqnarray}
where $G_{EE}^{>,ij}(q_{0})=\int\frac{{\rm d}^{3}\boldsymbol{q}}{(2\pi)^{3}}G_{EE}^{>,ij}(q)$
and $\Delta E=E_{n}-E_{m}$. 
In the slowly rotating case, rotation-induced splittings of otherwise degenerate magnetic sublevels
may be smaller than the detector's energy resolution.
These sublevels are then experimentally unresolved,
but their transition rates can still depend on the dipole matrix elements.
Such rate ratios can be modified by rotation through
the generalized KMS relation in Eq.~(\ref{eq:VF-KMS}),
\begin{eqnarray}
\frac{\Gamma_{n\to m}(\Delta E)}{\Gamma_{m\to n}(\Delta E)} & = & \frac{d_{mn}^{i}d_{mn}^{j*}G_{EE}^{>,ij}(-\Delta E)}{d_{mn}^{i}d_{mn}^{j*}G_{EE}^{>,ji}(\Delta E)}=e^{-\Delta E/T}\left(1+\frac{\omega^{k}}{T}\frac{i\epsilon^{ijk}d_{mn}^{i}d_{mn}^{j*}}{|\boldsymbol{d}_{mn}|^{2}}+\mathcal{O}(\mathcal{\omega}^2)\right) ,\label{eq:SM-Modified-DB}
\end{eqnarray}
where we introduce the transition matrix elements $d_{mn}^{i}=\langle m|\hat{d}^{i}|n\rangle$ and
$d_{mn}^{j,*}=\langle n|\hat{d}^{\dagger,j}|m\rangle$. 

For comparison, we study the modified detailed-balance for this specific
dipole interaction using the channel-resolved KMS relation 
$\frac{\Gamma_{n\to m}^{(\Delta l)}}{\Gamma_{m\to n}^{(-\Delta l)}}=e^{-\Delta E/T}e^{\omega\Delta l/T}$ \citep{Alicki:2017bgh}.
The E1 selection rule gives $\Delta l=l_{1}-l_{2}=0,\pm1$, 
so in the small-vorticity limit the spin-inclusive rate ratio (\ref{eq:SM-Modified-DB}) becomes
\begin{eqnarray}
\frac{\Gamma_{n\to m}(\Delta E)}{\Gamma_{m\to n}(\Delta E)} & = & \frac{\sum_{\Delta l}\Gamma_{n\to m}^{(\Delta l)}}{\sum_{\Delta l}\Gamma_{m\to n}^{(-\Delta l)}}= e^{-\Delta E/T}\left(1+\frac{\omega}{T}\frac{\Gamma_{m\to n}^{(\Delta l=1)}-\Gamma_{m\to n}^{(\Delta l=-1)}}{\Gamma_{m\to n}(\Delta E)}+\mathcal{O}(\mathcal{\omega}^2)\right).\label{eq:SM-Modified-DB-2}
\end{eqnarray}
Identifying $\Gamma_{m\to n}^{\Delta l=1}-\Gamma_{m\to n}^{\Delta l=-1}=i(d_{mn}^{x}d_{mn}^{y*}-d_{mn}^{y}d_{mn}^{x*})\rho(\Delta E)$ where $\rho(\Delta E)$ is the density of states, 
we recover the channel-resolved result Eq.~(\ref{eq:SM-Modified-DB-2})  from Eq.~(\ref{eq:SM-Modified-DB}) when $\boldsymbol{\omega}=\omega\hat{\boldsymbol{z}}$ is chosen. 

We remark that the spin-channel-resolved KMS relation in Ref.~\citep{Alicki:2017bgh}
is not sufficient to determine the relation between the vector-field Wightman function $G_{EE}^{>,ij}$
and the spectral function, and thus environment-dependent factors,
such as the density of states $\rho(E)$, cannot be factored out of Eq.~(\ref{eq:SM-Modified-DB-2}).
Another important difference is that the KMS relation in Ref.~\citep{Alicki:2017bgh}
applies only to scalar fields and misses the nontrivial contribution of spin angular momentum
to the vector-force correlator.
Another recent related work \citep{Salvio:2026ewl}
discusses the rotating KMS relation in the free case.
The resulting relation cannot be applied here to general environment correlations.
A spinful particle moving in the QGP acquires a continuous spin distribution and energy spectrum
due to its scattering-state nature.  In such a case, a detailed-balance relation similar to Eq.~(\ref{eq:SM-Modified-DB})
is presented in Eq.~(\ref{eq:Modified-detailed-balance}) for a general form of system-environment coupling,
whereas deriving it using a channel-resolved KMS relation is challenging.

\end{document}